\documentclass[twocolumn,superscriptaddress,nobalancelastpage,10pt,pra,letter]{revtex4}

\usepackage{amsmath}
\usepackage{amsthm}
\usepackage{amssymb}
\usepackage{amsfonts}
\usepackage{graphicx}
\usepackage{wasysym}
\usepackage{mathrsfs}
\usepackage{yfonts}
\usepackage{bbold}
\usepackage{verbatim}
\usepackage{color}
\usepackage{soul} 
\usepackage{lipsum}

\makeatletter
\newcommand*{\rom}[1]{\expandafter\@slowromancap\romannumeral #1@}
\makeatother

\begin{document}


\title{Position Correlation Enabled Quantum Imaging with Undetected Photons}

\author{Balakrishnan Viswanathan}
\affiliation{Department of Physics, Oklahoma State University, Stillwater, Oklahoma, USA}
\author{Gabriela Barreto Lemos} 
\affiliation{Physics Department, University of Massachusetts Boston,100 Morrissey Boulevard, Boston MA 02125, USA.}
\author{Mayukh Lahiri}
\email{mlahiri@okstate.edu} \affiliation{Department of Physics, Oklahoma State University, Stillwater, Oklahoma, USA}

\begin{abstract}
Quantum imaging with undetected photons (QIUP) is a unique imaging technique that does not require the detection of the light used for illuminating the object. The technique requires a correlated pair of photons. In the existing implementations of QIUP, the imaging is enabled by the momentum correlation between the twin photons. We investigate the complementary scenario in which the imaging is instead enabled by the position correlation between the two photons. We present a general theory and show that the properties of the images obtained in these two cases are significantly distinct.

\end{abstract}

\maketitle

\noindent

The field of quantum imaging has developed rapidly in recent decades \cite{moreau2019imaging}. One of the greatest achievements in this field is the discovery of new imaging techniques which are inspired by quantum physics.
\emph{Quantum imaging with undetected photons} (QIUP) is such a technique and its differentiating feature is that it does not  require the detection of the photon probing the object \cite{lemos_quantum_2014,lahiri2015theory}. This allows for probing samples at wavelengths for which appropriate cameras and detectors are not available. 
\par
QIUP requires a correlated photon pair (twin photons) produced in quantum superposition. One of the photons is used to probe the object (or sample) and the other photon is detected. The image is obtained through single-photon interference of the photon that never interacted with the object. No coincidence measurement or post-selection is used to obtain the image and this is one of the main distinctions of this imaging scheme with respect to other quantum imaging schemes, including ghost imaging \cite{pittman1995optical,gatti2008quantumimaging,aspden2013epr}. The QIUP is of particular interest when the illumination wavelength is one for which cameras and detectors are inadequate, as it allows the transfer of image information to a very different wavelength for detection. The imaging technique has recently been applied to microscopy \cite{Kviatkovskyeabd0264,paterova2020quantum} and also to image with intense beams \cite{cardoso}. 
\par
In the existing implementations of QIUP, both the object and the camera are placed in the far field relative to the twin photon source. Because of this, the imaging is enabled by the momentum correlation between the photon pair \cite{hochrainer2017quantifying,lahiri2017twin} and a recent study shows that the resolution is also determined by the momentum correlation \cite{fuenzalida2020resolution}. Furthermore, in this case, the image magnification depends on the ratio of wavelengths of the twin photons \cite{lemos_quantum_2014,lahiri2015theory}. 
\par
Twin photons are, in general, spatially entangled, i.e., in addition to the momentum correlation, they also exhibit the position correlation \cite{walborn2010spatial}. Therefore, it remains an open question: can the position correlation between the twin photons play a role in QIUP? Here, we answer this question by considering a scenario in which both the object and the camera are placed in the near field relative to the source of the twin photon. We present a rigorous theoretical analysis and show that the near-field imaging scheme has two striking differences from the far-field case: (1) the imaging is enabled by the position correlation of the photon pair instead of the momentum correlation, and (2) the image magnification does \emph{not} explicitly depend on the wavelengths of the twin photons. 
\par
Following standard terminology, we call two photons belonging to a pair signal ($S$) and idler ($I$). Throughout the analysis we assume that photons propagate as paraxial beams and are always incident normally on both the object and the detector. Under these assumptions, the two-photon quantum state can be written as (see, for example, \cite{walborn2010spatial})
\begin{equation} \label{state-spdc}
|\psi\rangle = \int d\textbf{q}_{s}  \ d\textbf{q}_{I} \ C(\textbf{q}_{s}, \textbf{q}_{I}) |\textbf{q}_{s}\rangle_{s} |\textbf{q}_{I}\rangle_{I},
\end{equation}
where $|\textbf{q}_{s}\rangle_{s} \equiv \hat{a}_{s}^{\dagger}(\textbf{q}_{s})|vac\rangle$ denotes a signal photon Fock state labeled by the transverse component $\textbf{q}_{s}$ of the wavevector $\textbf{k}_{s}$. Similarly, $|\textbf{q}_{I}\rangle_{I}\equiv \hat{a}_{I}^{\dagger}(\textbf{q}_{I})|vac\rangle $ denotes an idler photon Fock state labeled by the transverse component $\textbf{q}_{I}$ of the wavevector $\textbf{k}_{I}$. The complex quantity $ C(\textbf{q}_{S}, \textbf{q}_{I})$ ensures that $|\psi\rangle$ is normalized, i.e.,
\begin{equation} \label{C-fn-norm}
\int d\textbf{q}_{s}  \ d\textbf{q}_{I} \ |C(\textbf{q}_{s}, \textbf{q}_{I})|^{2} = 1.
\end{equation}  
\par
The joint probability density of detecting the signal and the idler photons at positions (transverse coordinates) $\boldsymbol{\rho}_{s}$ and $\boldsymbol{\rho}_{I}$, respectively, on the source plane is given by \cite{walborn2010spatial}
\begin{equation} \label{prob-pos}
P(\boldsymbol{\rho}_{s},\boldsymbol{\rho}_{I}) = \left|\int d \textbf{q}_{s} \ d \textbf{q}_{I}\  C(\textbf{q}_{s}, \textbf{q}_{I}) \ e^{i (\textbf{q}_{s}.\boldsymbol{\rho}_{s} + \textbf{q}_{I}.\boldsymbol{\rho}_{I})}\right|^{2}.
\end{equation}
The position correlation between the two photons is governed by this joint probability density. If $P(\boldsymbol{\rho}_{s},\boldsymbol{\rho}_{I})$ can be expressed as a product of a function of $\boldsymbol{\rho}_{s}$ and a function of $\boldsymbol{\rho}_{I}$, there is no position correlation. In the other extreme case, when the positions of the two photons are maximally correlated, the joint probability density is proportional to a delta function. 
\par
\begin{figure}
	\centering 
	\includegraphics[width=0.99\linewidth]{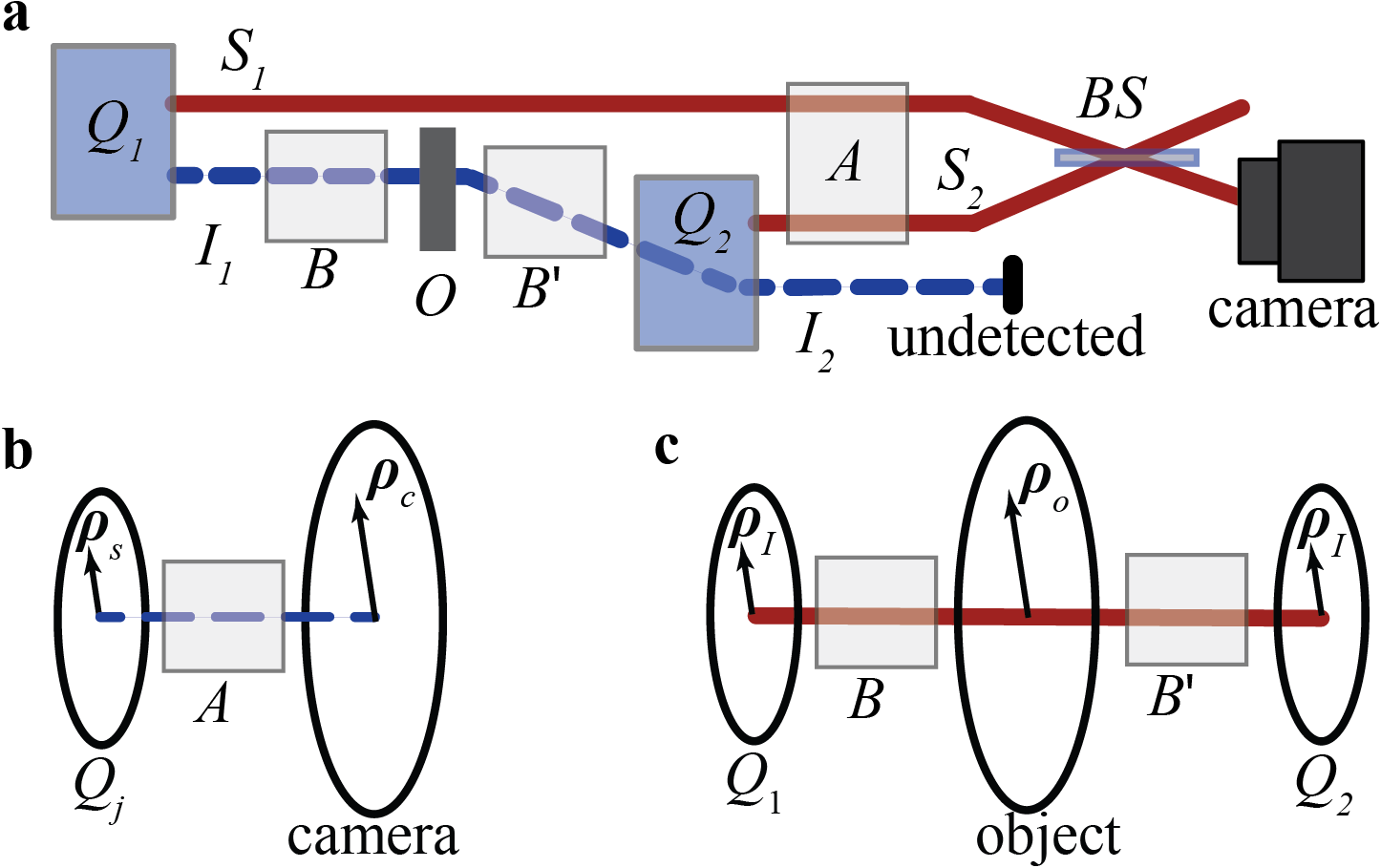}
	\caption{\textbf{a}, Schematics of the imaging scheme. Two identical twin photon sources, $Q_1$ and $Q_2$, can emit non-degenerate photon pairs (signal and idler) into beams $(S_{1},I_{1})$ and $(S_{2},I_{2})$. An imaging system, $B$ images the idler field at $Q_1$ onto the object, $O$, with magnification $M_I$. Another imaging system $B'$ images the idler field at the object onto $Q_2$ with magnification $1/M_I$. Idler beams $I_1$ and $I_2$ (dashed lines) are perfectly aligned and never detected. Signal beams $S_1$ and $S_{2}$ (solid lines) are superposed by a $50:50$ beam-splitter ($BS$) and projected onto a camera. An imaging system $A$ ensures that the signal field at the sources is imaged onto the camera with magnification $M_S$. The image of $O$ is obtained from the single-photon interference patterns observed at the camera without any coincidence measurement or postselection. \textbf{b}, A point $\boldsymbol{\rho}_{s}$ located on $Q_j~(j=1,2)$ is mapped on a point $\boldsymbol{\rho}_{c}=M_s\boldsymbol{\rho}_{s}$ on the camera by imaging system $A$. \textbf{c}, Due to the presence of imaging systems $B$ and $B'$, a point $\boldsymbol{\rho}_{I}$ on $Q_1$ is mapped onto a point $\boldsymbol{\rho}_{o}=M_I\boldsymbol{\rho}_{I}$ on the object and then again at $\boldsymbol{\rho}_{I}$ on $Q_2$.}
	\label{fig:img-schm}
\end{figure}

The imaging scheme is illustrated in Fig. \ref{fig:img-schm}a. There are two sources, $Q_1$ and $Q_2$, each of which can emit a photon pair. $Q_1$ emits the signal and idler photons into beams $S_1$ and $I_1$, respectively. Likewise, $S_2$ and $I_2$ represent the beams into which the signal and idler photons are emitted by $Q_2$. Usually, the sources are nonlinear crystals that produce photon pairs by spontaneous-parametric down conversion (SPDC) when weakly pumped by laser light. Our theoretical analysis is, however, valid for any source that may produce a spatially correlated photon pair. 
\par
The two signal beams ($S_{1}$ and $S_{2}$) are superposed by a $50:50$ beamsplitter (BS) and one of the outputs of BS is detected by a camera. An imaging system ($A$) with magnification $M_S$ ensures that the signal field at the sources is imaged onto the camera (Fig. \ref{fig:img-schm}a,b). (Since the sources have much smaller size compared to the path lengths in the interferometer, they can be assumed to be planar.) The signal photon \emph{never} interacts with the object. 
\par
The beam $I_1$ from source $Q_1$ illuminates the object, passes through source $Q_2$, and gets perfectly aligned with beam $I_2$. An imaging system, $B$, is placed between the source $Q_1$ and the object ($O$) in the beam $I_1$ such that the idler field at $Q_1$ is imaged onto the object with magnification $M_I$. Another imaging system, $B'$, images the idler field at the object onto source $Q_2$ with magnification $1/M_I$ (i.e., demagnified by the equal amount). These two imaging systems also ensure that $Q_2$ lies on the image plane of $Q_1$. For simplicity, we have assumed that magnifications of $B$ and $B'$ have the same sign. In order to obtain the best possible alignment of beams $I_1$ and $I_2$, it is absolutely essential that the magnitude of the total magnification due to the combined effect of $B$ and $B'$ is 1. We now show how the information of the object appears in the interference pattern recorded in the camera and how to construct the image from the interference pattern. 
\par
In QIUP, the two sources ($Q_1$ and $Q_2$) almost never emit simultaneously and almost never produce more than two photons individually, i.e., more than two photons are never simultaneously present in the system. In practice, such a situation is realized by pumping the nonlinear crystals very weakly such that the probability of generating more than one photon pair is negligible. Furthermore, the two sources emit coherently. In practice, this situation is realized by ensuring that the nonlinear crystals are pumped by mutually coherent laser beams. Under these circumstances, the quantum state of light generated by the two sources is given by the superposition of the states generated by them individually, i.e., by
\begin{align} \label{biphoton-state}
&|\Psi\rangle = \alpha_{1} \ \int d\textbf{q}_{I_{1}} \  d\textbf{q}_{s_{1}} \ C(\textbf{q}_{s_{1}}, \textbf{q}_{I_{1}}) \ \hat{a}^{\dagger}_{I_{1}}(\textbf{q}_{I_{1}}) \ \hat{a}^{\dagger}_{s_{1}}(\textbf{q}_{s_{1}}) \ |vac \rangle \cr
&+ \alpha_{2} \ \int d\textbf{q}_{I_{2}}\ d\textbf{q}_{s_{2}} \ C(\textbf{q}_{s_{2}}, \textbf{q}_{I_{2}}) \ \hat{a}^{\dagger}_{I_{2}}(\textbf{q}_{I_{2}}) \ \hat{a}^{\dagger}_{s_{2}}(\textbf{q}_{s_{2}}) \ |vac \rangle,
\end{align}
where $\alpha_{1}$ and $\alpha_{2}$ are complex numbers satisfying the condition $|\alpha_{1}|^{2} + |\alpha_{2}|^{2} = 1$, and $|vac \rangle$ represents the vacuum state.
\par
The effect of the object on the idler field is practically equivalent to that of a beamsplitter with one input. Taking the imaging systems $B$ and $B'$ into account, we can therefore relate the quantum field, $\hat{E}_{I_{2}}^{(+)}(\boldsymbol{\rho}_{I})$, associated with the idler photon at $Q_2$ to the quantum field, $\hat{E}_{I_{1}}^{(+)}(\boldsymbol{\rho}_{I})$, associated with the idler photon at $Q_1$ in the following way: 
\begin{align}\label{align-pos}
\hat{E}_{I_{2}}^{(+)}(\boldsymbol{\rho}_{I}) = e^{i \phi_{I}'(\boldsymbol{\rho}_{I})}\big[& e^{i \phi_{I}(\boldsymbol{M_I\rho}_{I})}  \ T (M_{I} \boldsymbol{\rho}_{I}) \hat{E}_{I_{1}}^{(+)}(\boldsymbol{\rho}_{I})  \nonumber \\ & + \  R (M_{I} \boldsymbol{\rho}_{I}) \hat{E}_{0}^{(+)}(\boldsymbol{\rho}_{I}) \big],
\end{align}
where $\hat{E}_{0}^{(+)}(\boldsymbol{\rho}_{I})$ is the corresponding vacuum field, $\phi_{I}(M_I\boldsymbol{\rho}_{I})$ and $\phi_{I}'(\boldsymbol{\rho}_{I})$ are the phase introduced by the imaging systems $B$ and $B'$, respectively, $T(M_{I} \boldsymbol{\rho}_{I})$ is the complex amplitude transmission coefficient of the object at a point $\boldsymbol{\rho}_{o} \equiv M_{I} \boldsymbol{\rho}_{I}$ for normal incidence, and $R(M_{I} \boldsymbol{\rho}_{I})=\sqrt{1-|T(M_{I} \boldsymbol{\rho}_{I})|^2}$. We stress that the effect of stimulated emission due to this alignment is negligible \cite{zou1991induced,wiseman2000induced,kolobov2017controlling,lahiri2019nonclassical}.
\par
We now represent the idler field at the sources ($z=0$) in terms of its angular spectrum (\cite{monken1998transfer}; see also \cite{MW}, Sec. 3.2). Since $Q_1$ is imaged onto $Q_2$, we can write the following expression without any loss of generality:
\begin{equation}\label{e-idler-angs}
\hat{E}_{I_{j}}^{(+)}(\boldsymbol{\rho}_{I}) = \int d \textbf{q}_{I_{j}} \ \hat{a}_{I_{j}}(\textbf{q}_{I}) \ e^{i \textbf{q}_{I_{j}}\cdot \boldsymbol{\rho}_{I}} ,
\end{equation}
where $j=1,2$ correspond to the sources $Q_1$ and $Q_2$, respectively. Substituting from Eq. (\ref{e-idler-angs}) into Eq. (\ref{align-pos}) and applying the convolution theorem \cite{weber2005mathematical}, we find that
\begin{align}\label{align-idler op}
\hat{a}_{I_{2}}(\textbf{q}_{I}) = \int &d \textbf{q}_{I}^{'} \frac{1}{M_{I}^2} \big[ \widetilde{T}'\left(\frac{\textbf{q}_{I}- \textbf{q}_{I}^{'}}{M_{I}} \right) \ \hat{a}_{I_{1}}(\textbf{q}_{I}^{'})  \nonumber \\
&+ \widetilde{R}'\left(\frac{\textbf{q}_{I}- \textbf{q}_{I}^{'} }{M_{I}} \right)  \ \hat{a}_{0}(\textbf{q}_{I}^{'}) \big],
\end{align}
where $\widetilde{T}'(\textbf{q}_{I}/M_{I})$ and $\widetilde{R}'(\textbf{q}_{I}/M_{I})$ are the Fourier transforms of $\text{exp}[i\{\phi_{I}(M_I\boldsymbol{\rho}_{I})+\phi_{I}'(\boldsymbol{\rho}_{I})\}] T(M_{I} \boldsymbol{\rho}_{I})$ and $\text{exp}[i\phi_{I}'(\boldsymbol{\rho}_{I})]R(M_{I} \boldsymbol{\rho}_{I})$, respectively. From Eqs. (\ref{biphoton-state}) and (\ref{align-idler op}), we find that the quantum state of light generated by the system is given by
\begin{align} \label{biphoton-state-align}
&|\psi\rangle = \alpha_{1} \int d \textbf{q}_{I_{1}} \ d \textbf{q}_{s_{1}} \ C(\textbf{q}_{I_{1}}, \textbf{q}_{s_{1}}) \   |\textbf{q}_{I_{1}}\rangle_{I_{1}}  |\textbf{q}_{s_{1}}\rangle_{s_{1}} \nonumber \\
&+ \alpha_{2}  \int d \textbf{q}_{I_{2}}\ d \textbf{q}_{s_{2}} \ d \textbf{q}_{I}^{'}  \ C(\textbf{q}_{I_{2}}, \textbf{q}_{s_{2}}) \nonumber \\
& \qquad \quad \times  \frac{1}{M_{I}^2}\Big[\widetilde{T}'^{*}\left(\frac{\textbf{q}_{I_{2}}- \textbf{q}_{I}^{'}}{M_{I}} \right) \    |\textbf{q}^{'}_{I}\rangle_{I_{1}} \nonumber \\
&\qquad \qquad + \widetilde{R}'^{*}\left(\frac{\textbf{q}_{I_{2}}- \textbf{q}_{I}^{'} }{M_{I}} \right) \   |\textbf{q}^{'}_{I}\rangle_{0} \Big]  |\textbf{q}_{s_{2}}\rangle_{s_{2}},   
\end{align}
where $|\textbf{q}\rangle_{0}=\hat{a}_{0}^{\dag}(\textbf{q})|vac \rangle$. 
\par
We represent the signal field at each source ($z=0$) by its angular spectrum. Since the signal beams are superposed by a $50:50$ beamsplitter ($BS$) and both sources are imaged onto the camera with magnification $M_{s}$, the positive frequency part of the total signal field at a point $\boldsymbol{\rho}_{c} \equiv M_{s}\boldsymbol{\rho}_{s}$ on the camera is given by
\begin{align} \label{detection operator}
\hat{E}_{s}^{(+)}(\boldsymbol{\rho}_{c}) \propto & \int d\textbf{q}_{s} \Big[ \ \hat{a}_{s_{1}}(\textbf{q}_{s}) \nonumber \\& \qquad + i \ e^{[i \phi_{s0}+\phi_s(\boldsymbol{\rho}_{c})]} \ \hat{a}_{s_{2}}(\textbf{q}_{s})  \Big] e^{i \ \textbf{q}_{s} \cdot \boldsymbol{\rho}_{c}/M_s},
\end{align}
where the phase difference between the two signal fields is written as a combination of $\phi_{s0}$ and $\phi_s(\boldsymbol{\rho}_{c})$; the former is a spatially independent phase that can be varied to obtain interference patterns and the latter is a spatially dependent phase that may arise due to the presence of imaging system $A$.
\par
The photon counting rate at a point $\boldsymbol{\rho}_{c}$ on the camera is determined by the standard formula $\mathcal{R}(\boldsymbol{\rho}_{c}) \propto \langle\psi|\hat{E}_{s}^{(-)}(\boldsymbol{\rho}_{c}) \ \hat{E}_{s}^{(+)}(\boldsymbol{\rho}_{c})|\psi\rangle,$
where $ \hat{E}_{s}^{(-)}(\boldsymbol{\rho}_{c}) = [\hat{E}_{s}^{(+)}(\boldsymbol{\rho}_{c})]^{\dagger}$. Using Eqs.  (\ref{prob-pos}), (\ref{biphoton-state-align}), and (\ref{detection operator}) we find that
\begin{align} \label{photon count-final-expr}
&\mathcal{R}(\boldsymbol{\rho}_{c}) \nonumber \\
& \propto  \int d \boldsymbol{\rho}_{o} \ P\left(\frac{\boldsymbol{\rho}_{c}}{M_s}, \frac{\boldsymbol{\rho}_{o}}{M_I} \right) \big( |\alpha_{1}|^{2} + |\alpha_{2}|^{2} + 2 |\alpha_{1}||\alpha_{2}||T(\boldsymbol{\rho}_{o})|   \nonumber \\
& \quad \times \text{cos}[\phi_{in}+\phi_s(\boldsymbol{\rho}_{c}) -\phi_{I}(\boldsymbol{\rho}_{o})-\phi_{I}'(\frac{\boldsymbol{\rho}_{o}}{M_I})- \phi_{T}(\boldsymbol{\rho}_{o}) ]\big),
\end{align} 
where $\phi_{in}=\phi_{s0} + \text{arg}\{\alpha_{2}\} - \text{arg}\{\alpha_{1}\}$. 
\par
It follows from Eq. (\ref{photon count-final-expr}) that both the magnitude $(|T(\boldsymbol{\rho}_{o})|)$ and phase $(\phi_{T}(\boldsymbol{\rho}_{o}))$ of the transmission coefficient of the object appear in the photon counting rate (intensity) observed on the camera, even though the photons interacting with the object were never detected by the camera. Equation (\ref{photon count-final-expr}) also shows that information about a range of points on the object plane, averaged by the joint probability distribution $P$, appears at a single point on the camera. Since this probability distribution characterized the position correlation between the twin photons (see Eq. (\ref{prob-pos})), it becomes evident that the position correlation plays the key role in the image formation in this case.
\par
We now illustrate the image formation by considering the case in which the positions of the photon pair are maximally correlated, i.e.,
\begin{equation} \label{delta corr}
P\left(\frac{\boldsymbol{\rho}_{c}}{M_s}, \frac{\boldsymbol{\rho}_{o}}{M_I} \right)=P(\boldsymbol{\rho}_{s}, \boldsymbol{\rho}_{I}) \propto \delta(\boldsymbol{\rho}_{s}-\eta \boldsymbol{\rho}_{I}),
\end{equation}
where $\eta$ is a dimensionless scalar parameter and we recall that $\boldsymbol{\rho}_{c}=M_s\boldsymbol{\rho}_{s}$ and $\boldsymbol{\rho}_{o}=M_I\boldsymbol{\rho}_{I}$. On substituting from Eq. (\ref{delta corr}) into Eq. (\ref{photon count-final-expr}) and setting $|\alpha_{1}| = |\alpha_{2}| = 1/\sqrt{2}$ for simplicity, we find that
\begin{align} \label{Photon count-delta}
&\mathcal{R}(\boldsymbol{\rho}_{c})\propto 1 + \left|T \left(\boldsymbol{\rho}_{o} \right) \right| \nonumber \\ & \times \cos \big[\phi_{in} +\phi_s(\boldsymbol{\rho}_{c})- \phi_{I}\left(\boldsymbol{\rho}_{o} \right)- \phi_{I}' \left(\frac{\boldsymbol{\rho}_{o}}{M_{I}} \right) - \phi_{T}\left( \boldsymbol{\rho}_{o} \big)\right]
\end{align}
and $\boldsymbol{\rho}_{c}=\eta (M_{s}/M_{I}) \boldsymbol{\rho}_{o}$. It is evident that if $\phi_{in}$ is varied, the photon counting rate (intensity) at each point on the camera varies sinusoidally, i.e., a single-photon interference pattern is generated at each point on the camera. We also notice that in this case information about a single point on the object plane appears at a single point on the camera.
\par
If we determine the visibility of the single-photon interference pattern at a point $(\boldsymbol{\rho}_{c})$ on the camera, we find that
\begin{align} \label{abs.object}
\mathcal{V}(\boldsymbol{\rho}_{c}) &\equiv \frac{\mathcal{R}_{max}(\boldsymbol{\rho}_{c}) - \mathcal{R}_{min}(\boldsymbol{\rho}_{c})}{\mathcal{R}_{max}(\boldsymbol{\rho}_{c}) + \mathcal{R}_{min}(\boldsymbol{\rho}_{c})} =\left|T(\boldsymbol{\rho}_{o})\right|, 
\end{align}
where we have used the relation between $\boldsymbol{\rho}_{c}$ and $\boldsymbol{\rho}_{o}$ given below Eq. (\ref{Photon count-delta}). For an absorptive object, we can set $\phi_{T}(\boldsymbol{\rho}_{o})=0$, i.e., $T(\boldsymbol{\rho}_{o})=|T(\boldsymbol{\rho}_{o})|$. Therefore, the image of an absorptive object is obtained from the visibility of the interference patterns observed on the camera.
\begin{figure}[htbp]
	\centering
	\includegraphics[width=1.00\linewidth]{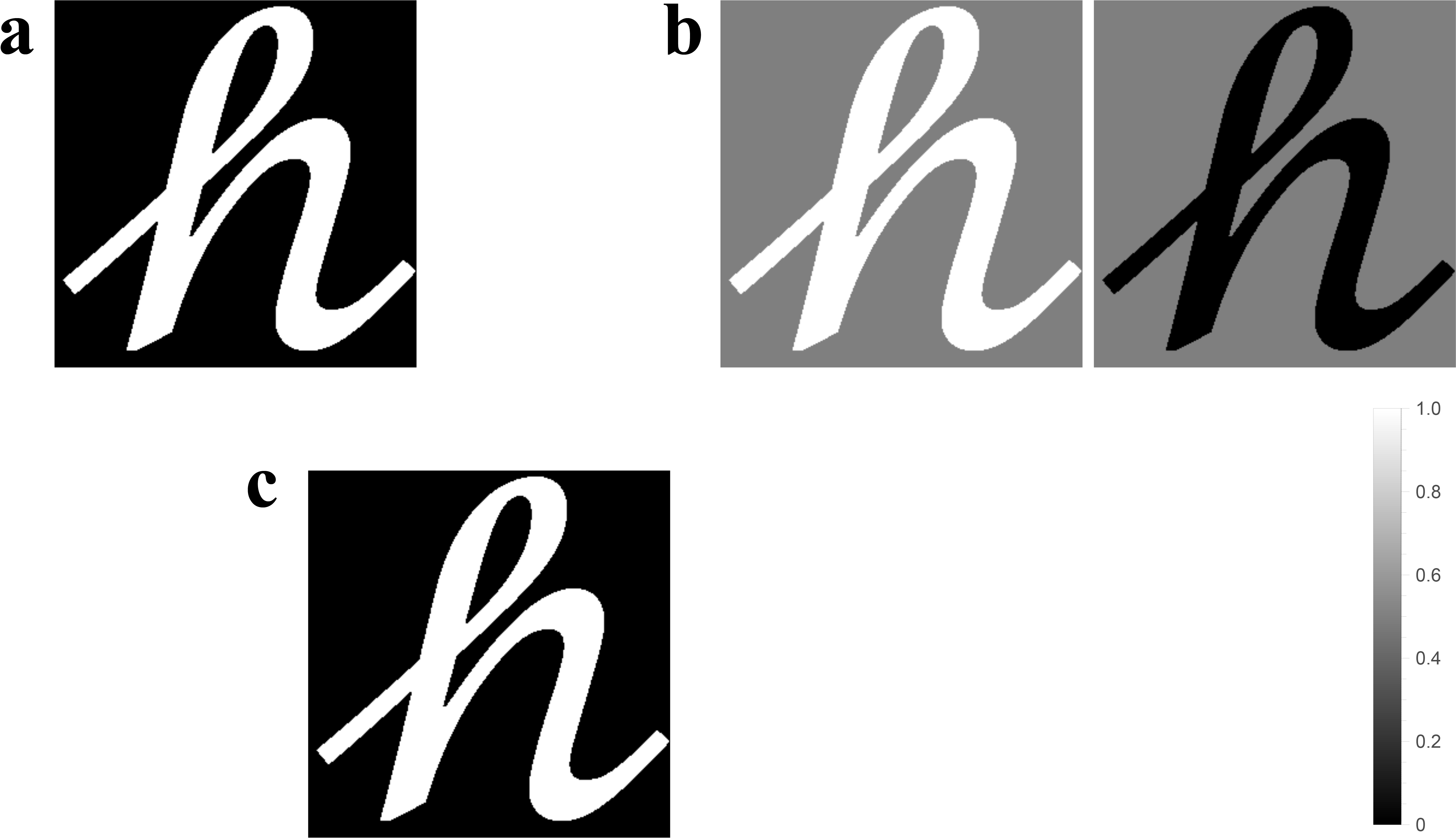}
	\caption{Imaging an absorptive object. \textbf{a}, The object: the amplitude transmission coefficient is $0$ and $1$ in the black and white regions, respectively. \textbf{b}, Normalized single-photon counting rates (intensities) for constructive (left) and destructive (right) interference show that the information of the object is present in the interference patterns observed on the camera. The phases introduced by the imaging systems are assumed to have no spatial dependence. \textbf{c}, The visibility measured at each point on the camera gives the image. The magnification is 1.} \label{fig:ImagingFig}
\end{figure}
\par
From the relation between $\boldsymbol{\rho}_{c}$ and $\boldsymbol{\rho}_{o}$ given below Eq. (\ref{Photon count-delta}), we also find that the image magnification is given by
\begin{equation} \label{image-mag}
M = \eta \frac{M_{s}}{M_{I}},
\end{equation}
where $M_{s}$ and $M_{I}$ are the magnifications of the imaging systems, $A$ and $B$, placed on signal and idler paths, respectively, and $\eta$ is a dimensionless scalar quantity introduced in Eq. (\ref{delta corr}). Generalizing the result shown in Ref. \cite{walborn2010spatial}, Eq. (59), to the  non-degenerate spontaneous parametric down-conversion [see, for example, Ref. \cite{hochrainer2017quantifying}, supporting information, Eq. (S3)] and using Eq. (\ref{prob-pos}), one can readily show that for standard non-degenerate SPDC processes $\eta = 1$, i.e., 
\begin{equation} \label{image-mag-2}
M = \frac{M_{s}}{M_{I}}.
\end{equation} 
Since signal and idler photons pass through independent imaging systems, magnifications $M_{I}$ and $M_{s}$ can be chosen independently.
Therefore, Eq. (\ref{image-mag-2}) shows that the image magnification does \emph{not} explicitly depend on wavelength. This is a striking difference from the case of far field imaging with undetected photons where the magnification must explicitly depend on the ratio of wavelengths of the twin photons. In the far field case, if one swaps the undetected and detected wavelengths while keeping the other parameters fixed, the magnification must change (for non-degenerate twin photons). However, in the near field case, the same action does not change the magnification.
\par
Figure \ref{fig:ImagingFig} illustrates image construction for an absorptive object with magnification $1$. The object is shown in Fig. \ref{fig:ImagingFig}(a). For simulating the photon-counting rate (intensity), we assumed the simple case in which the phase introduced by all the imaging systems are spatially independent. In Fig. \ref{fig:ImagingFig}(b), normalized photon counting rates (intensities) for constructive (left) and destructive interference are shown. It is evident that the information of the object is present in the photon counting rates. Figure \ref{fig:ImagingFig}(c) shows the visibility measured at each point on the camera. Clearly, the image of the absorptive object is given by the visibility map. 
\par
In the case of a purely phase object, we can set $|T(\boldsymbol{\rho}_{o})| = 1$, i.e., $T(\boldsymbol{\rho}_{o}) = \text{exp}[i \phi_{T}(\boldsymbol{\rho}_{o})]$. Equation (\ref{Photon count-delta}) now reduces to
\begin{align} \label{phase object}
&\mathcal{R}(\boldsymbol{\rho}_{c}) \propto 1  \nonumber \\ &+  \cos \big[\phi_{in} +\phi_s(\boldsymbol{\rho}_{c})- \phi_{I}\left(\boldsymbol{\rho}_{o} \right)- \phi_{I}' \left(\frac{\boldsymbol{\rho}_{o}}{M_{I}} \right) - \phi_{T}\left( \boldsymbol{\rho}_{o} \right)\big].
\end{align}
Since the spatially dependent phases $\phi_s$, $\phi_{I}$, and $\phi_{I}'$ are introduced by the imaging systems, they are known quantities, e.g., for $4f$ imaging systems, $\phi_s$, $\phi_{I}$, and $\phi_{I}'$ can be treated as constants. Therefore, the phase of the amplitude transmission, $\phi_{T}(\boldsymbol{\rho}_{o})$, can also be determined using standard procedures (see, for example, \cite{mir2012quantitative}), i.e., a phase object can also be imaged using this scheme. For objects with relatively simple distribution of phase, such as the ones considered in Ref. \cite{lemos_quantum_2014}, the image can also be obtained by image subtraction. The magnification will once again be given by Eq. (\ref{image-mag}). 
\par
Finally, we would like to point out that the two twin photon sources, $Q_1$ and $Q_2$, do not need to be spatially separated. In fact, a single nonlinear crystal pumped by two mutually coherent laser beams from two sides is fully equivalent to two identical sources for imaging purpose. Pumping a single crystal from two sides was experimentally demonstrated in Ref. \cite{herzog1994frustrated} and its application for imaging was proposed in \cite{QIUP-patent}. The theory we have developed in this letter readily applies to such a configuration. It is also straightforward to extend the theory to cover reflective objects for which the spatially dependent reflection coefficient can be determined from the interference pattern in a similar manner. 
\par
In conclusion, we have explored a new aspect of quantum imaging with undetected photons. We have shown that when both the object and camera are placed in the near field relative to the twin photon sources, the imaging is enabled by the transverse position correlation between the photon pairs. We have also shown that the magnification of the image does not depend on the wavelengths of the photon pair. In contrast, far field quantum imaging with undetected photons relies on the transverse momentum correlations and the image magnification depends on the wavelengths of the photon pair \cite{lemos_quantum_2014,lahiri2015theory,lahiri2017twin,hochrainer2017quantifying,fuenzalida2020resolution}. Our analysis strongly suggests that in the near field case, the resolution will be determined by this position correlation. Since study of resolution require separate attention, we do not discuss the topic in this letter. We expect that our work will provide deeper understanding of the imaging process paving the way for a resolution enhancement in quantum imaging with undetected photons. 
\par
\textbf{Acknowledgements.} B.V. and M.L. acknowledge support from College of Arts and Sciences and the Office of the Vice President for Research, Oklahoma State University.

\bibliography{img-references}

\end{document}